\documentclass[twocolumn,floats,superscriptaddress]{revtex4}
\usepackage{graphicx,epsfig}
\usepackage{times}
\usepackage{graphics,dcolumn,bm,float}
\usepackage{amssymb,amsmath,rotate,color}
\usepackage[normalem]{ulem}

\begin{document}
\title{Band-offset-induced lateral shift of valley electrons in ferromagnetic MoS$_2$/WS$_2$ planar heterojunctions}
\author{Hassan Ghadiri}
\affiliation{Department of Physics, North Tehran Branch, Islamic
Azad University, 16511-53311, Tehran, Iran}
\author{Alireza Saffarzadeh}\email{asaffarz@sfu.ca}
\affiliation{Department of Physics, Payame Noor University, P.O.
Box 19395-3697 Tehran, Iran} \affiliation{Department of Physics,
Simon Fraser University, Burnaby, British Columbia V5A 1S6,
Canada}
\date{\today}

\begin{abstract}
Low-energy coherent transport and Goos-H\"{a}nchen (GH) lateral
shift of valley electrons in planar heterojunctions composed of
normal MoS$_2$ and ferromagnetic WS$_2$ monolayers are
theoretically investigated. Two types of heterojunctions in the
forms of WS$_2$/MoS$_2$/WS$_2$ (type-A) and MoS$_2$/WS$_2$/MoS$_2$
(type-B) with incident electrons in MoS$_2$ region are considered
in which the lateral shift of electrons is induced by band
alignments of the two constituent semiconductors. It is shown that
the type-A heterojunction can act as an electron waveguide due to
electron confinement between the two WS$_2$/MoS$_2$ interfaces
which cause the incident electrons with an appropriate incidence
angle to propagate along the interfaces. In this case the spin-
and valley-dependent GH shifts of totally reflected electrons from
the interface lead to separated electrons with distinct
spin-valley indexes after traveling a sufficiently long distance.
In type-B heterojunction, however, transmission resonances occur
for incident electron beams passing through the structure, and
large spin- and valley-dependent lateral shift values in
propagating states can be achieved. Consequently, the transmitted
electrons are spatially well-separated into electrons with
distinct spin-valley indexes. Our findings reveal that the planar
heterojunctions of transition metal dichalcogenides can be
utilized as spin-valley beam filter and/or splitter without
external gating.

\end{abstract}
\maketitle

\section{Introduction}
Monolayers of transition metal dichalcogenides (TMDs) MX$_2$ (M =
Mo, W; X = S, Se, Te) with large intrinsic band gap have unique
electronics and optoelectronics properties which show their
potential applications for creating new kinds of nanodevices
\cite{Radisavljevic,Perkins,Yin,Sundaram,Xiao}. Intrinsic strong
spin-orbit coupling and the absence of inversion symmetry in these
monolayers lead to a giant spin splitting at the K point of their
hexagonal Brillouin zone. Moreover, due to a large valley
separation in their momentum space, the valley index is regarded
as a discrete degree of freedom for low-energy carriers.
Therefore, the valley index, like charge and spin, can be used to
encode information in TMD monolayers \cite{Wu1}.

Furthermore, heterostructures of TMD monolayers exhibit junctions
with novel properties that are unobtainable from individual MX$_2$
monolayers and can be used as building blocks of optoelectronic
devices, such as light emitting diodes and photodetectors
\cite{Lopez,Radis}. Recently various in-plane heterostructures
such as MoS$_2$/WS$_2$ \cite{Gong,K-Chen}, MoS$_2$/MoSe$_2$ and
WS$_2$/WSe$_2$ \cite{Duan}, MoSe$_2$/WSe$_2$ \cite{Huang}, and
WSe$_2$/MoS$_2$ \cite{Li} have been successfully prepared and
microstructures and morphologies of these seamless and atomically
sharp planar heterojunctions have been characterized. Performance
and functionality of these structures are critically dependent on
the alignment of their energy bands. In this regard, theoretical
calculations based on first principals studies have been made to
determine the band-offset \cite{Kang} and other properties of TMD
lateral heterojunctions \cite{Peeters,An}.

It is known from optics that a beam of light that is totally
reflected from the interface between two media undergoes a lateral
displacement along the interface, known as Goos-H\"{a}nchen (GH)
shift \cite{Newton,Goos}. In GH effect the incident wave packet of
plane waves is reshaped by the interface due to a different phase
shift that each plane wave in the light wave packet experiences.
In this regard, the GH effect of light beams was observed in
several experiments (see Ref. [\onlinecite{OpticExpress}] and
references therein). In addition to this spatial shift which is
dependent on the polarization of incident beam, angular shifts in
reflection of light beam by an air-glass interface have also been
reported \cite{NaturePhotonics}.

The GH effect has been spread to various areas of physics
\cite{Briers,Huang2,Ignatovich,Zhou}, specially condensed matter
systems including semiconductors \cite{Wilson}, two-dimensional
(2D) materials \cite{Beenakker,Sun}, topological insulators
\cite{Kuai}, and spin waves \cite{Gruszecki}. Moreover, lateral
resonance shifts of transmitted electrons through semiconductor
quantum barriers and wells were studied when electron beams,
incident from outside of the well/barrier region, propagate
through the structures \cite{Chen2006,Chen2009}. Such lateral
shifts (displacements), like the GH effect originating from beam
reshaping are named Goos-H\"{a}nchen-like ( GHL) shifts
\cite{ChenEPJB} as well as the lateral shift at the transmitted
resonances in the literature \cite{Xiang}.

Many studies have been devoted to the investigation of GH and GHL
shifts in graphene-based structures, such as graphene $p$-$n$
junction \cite{Beenakker}, barriers \cite{ChenEPJB,Ghosh,Song},
and superlattices \cite{Chen2013} in both Klein tunneling
\cite{Klein} and classical motion regimes \cite{ChenEPJB}. Also,
it was shown that valley-dependent GH \cite{Wu} and GHL
\cite{Zhai} shifts can be produced by local strains on
single-layer graphene without requiring any external fields. In
addition, the GH effect of electrons has been studied in a p-n-p
junction of MoS$_2$ monolayer \cite{Sun} and it was shown that the
shift is spin- and valley-dependent, due to spin-valley coupling
in MoS$_2$ monolayer. Similar results have also been reported in
the GHL shift of electron beam transmitted through a ferromagnetic
silicene \cite{Azarova}. Moreover, in our recent work
\cite{Ghadiri}, GHL shift of both transmitted and reflected
electrons in a gated monolayer WS$_2$ was studied. Interestingly,
it was shown that in contrast to the transmitted beam, the GHL
shift of reflected electrons is not invariant under simultaneous
interchange of spin and valley indexes.

In most of the previous models the GH (GHL) shift of electrons is
induced by either an applied gate voltage or a uniaxial strain in
a specific region of the structure. In lateral TMD
heterojunctions, however, the band offset between two constituent
materials can generate a lateral shift which is controllable by
energy and incidence angle of electron beam. On the other hand,
placing a MX$_2$ monolayer on an insulating magnetic substrate can
make the material ferromagnetic. Therefore, the
band-offset-induced lateral shift and the proximity-induced
ferromagnetic order in MX$_2$ planar heterojuntions can lead to
novel device applications, such as spin-valley filters and/or
splitters which are potentially useful for valley-spintronics.

In this paper we study quantum transport and band-offset-induced
lateral shift of valley electrons in planar heterojunctions
composed of normal MoS$_2$ and ferromagnetic WS$_2$ monolayers.
These MoS$_2$/WS$_2$ lateral heterostructures with common sulphur
have a type-II band alignment and the valence (conduction) band of
WS$_2$ is 0.39 eV (0.35 eV) higher than that of MoS$_2$
\cite{Kang}. We show that an incident beam of electrons in the
MoS$_2$ region can be confined between two WS$_2$ monolayers,
depending on the Fermi energy and the incidence angle of
electrons. The confined electrons will be separated into electrons
with distinct spin and valley indexes after passing a sufficiently
long distance in the MoS$_2$ region acting as an electron
waveguide. On the other hand, in a heterojunction with two MoS$_2$
monolayers and a single-layer WS$_2$ in between, transmission
resonances and large lateral shifts can occur for incident
electron beams propagating through the structure. As a result, the
transmitted electron beams can be spatially well separated into
electrons with distinct spin and valley indexes.

The paper is organized as follows. In section II, we introduce our
model and formalism for calculation of spin-valley transport and
lateral shift values for two types of heterojunctions, named
type-A and type-B, ignoring the electron-electron, hole-hole, and
electron-hole interactions. By tuning our system parameters,
numerical results and discussions for both types of
heterojunctions are presented in Sec. III. The GH effect is
discussed in type-A heterojunction, whereas the GHL effect is
described in type-B heterojunction. A brief conclusion is given in
Sec. IV.

\section{Model and formalism}
We consider two types of lateral heterojunctions consisting of two
monolayers MoS$_2$ and WS$_2$, in the form of
WS$_{2}$/MoS$_{2}$/WS$_{2}$ (type-A) and
MoS$_{2}$/WS$_{2}$/MoS$_{2}$ (type-B) in $x-y$ plane, as shown in
Figs. 1(a) and 1(b), respectively. In type-A heterojunction, the
electron beam is incident in the central region $0\leq x\leq d$
(region 2), so that it can propagate along the $y$ direction with
translational invariance, whereas they are totally reflected from
MoS$_{2}$/WS$_{2}$ interfaces at $x=0$ and $x=d$, leading to
electron confinement along the $x$ direction. In such a case the
heterojunction can act as an electron waveguide (see Fig. 1(a)).
In the case of type-B heterojunction, however, the electron beam
is incident from $x<0$ (region 1) on the MoS$_{2}$/WS$_{2}$
interface at $x=0$ and partially reflected to the same region, and
partially transmitted into the region 3 at $x>d$ (see Fig. 1(b)).
The influence of an exchange field $\mathbf{h}=h\hat{z}$ induced
by magnetic proximity effect and originated from an insulating
ferromagnetic substrate is assumed on each WS$_2$ monolayer. In
fact the localized magnetic moments in the ferromagnetic insulator
induce an exchange field that acts as an effective Zeeman field on
electrons in the structure. This interaction is short ranged and
only the nearest layer of magnetic ions contributes in this field
\cite{Haugen}. Therefore, in Fig. 1, it is reasonable to assume
that the exchange field is confined in the WS$_2$ region and
neglect its influence on the MoS$_2$ regions. Such a
magnetic-exchange field which has also been studied in graphene
\cite{Yokoyama1,Haugen}, silicene \cite{Yokoyama2,Azarova}, and
MoS$_2$ \cite{Li1} increases the spin splitting of the valence and
conduction bands in the materials. Note that MoS$_2$ and WS$_2$
monolayers have the same crystal structure and the mismatch
between their lattice constants is less than 0.22\% \cite{Yoo}. On
the other hand, recent observations clearly show atomically clean
and sharp junction between WS$_2$ and MoS$_2$ along zigzag-edge
directions \cite{Yoo,Gong}. Therefore, the edge effects from
WS$_2$ region on the MoS$_2$ are ignored in this study.

Denoting the electron wavefunctions in the valence and conduction
bands of region $j$ ($=1,2,3$) as $\psi_{jv}$ and $\psi_{jc}$,
respectively, the low energy electrons with energy $E$ near the
valleys K ($\tau=1$) and K$^\prime (\tau=-1$) in the presence of
exchange field $h_j$ satisfy the following Dirac-like equation
\cite{Xiao}
\small
\begin{equation}\label{h1}
\left(\begin{array}{cc} E_{jc}-E-h_js_z & \tau
a_jt_jk_je^{-i\tau\theta
_j} \\
\tau a_jt_jk_je^{i\tau\theta_j} & E_{jv}+\tau s_z\lambda_j-E-h_js_z \\
\end{array}\right)
\left(\begin{array}{c}
\psi_{jc} \\
\psi_{jv} \\
\end{array}\right)=0\  ,
\end{equation}
\normalsize
where $E_{jc}$ $(E_{jv})$ is the energy of conduction
(valence) band minimum (maximum) in the absence of exchange field
and spin-orbit coupling, $a_j$ is the lattice constant, $t_j$ is
the effective hopping integral, and $2\lambda_j$ is the spin
splitting at the valence band edges due to the spin-orbit coupling
in $j$th region. Moreover, $s_z=+1(-1)$ is the spin of electron,
and $k_j$ and $\theta_j$ are the magnitude and angle (relative to
the $x$-axis) of electron wave vector, $\mathbf{k}_j$, in $j$th
region, respectively.

Solving Eq. (\ref{h1}), we obtain dispersion relation and
pseudospinor components as
\begin{eqnarray}\label{Ek}
\begin{array}{cc}
(2E-(E_{jc}+E_{jv})-\tau s_z\lambda_j+2h_js_z)^2 \\
-(E_{jc}-E_{jv}-\tau s_z\lambda_j)^2=(2a_jt_jk_j)^2 \  ,
\end{array}
\end{eqnarray}
and
\begin{eqnarray}\label{h2}
\left(\begin{array}{c} \psi_{jc} \\
\psi_{jv} \\
\end{array}\right)=\frac{1}{B_j}
\left(\begin{array}{c}
A_j \\
\tau a_jt_jk_je^{i\tau\theta_j} \\
\end{array}\right)
\mathrm{e}^{i(k_{jx}x+k_yy)}\  ,
\end{eqnarray}
where
$A_j=E-E_{jv}-\tau s_z\lambda_j+h_js_z$ and
$B_j=\sqrt{A_j^{2}+(a_jt_jk_j)^{2}}$ .

In the following subsections, we study the behavior of electron
beam by obtaining formulas for transmission probability $T$ and GH
(GHL) lateral shift of electrons propagating in type-A(B)
heterojunctions. We mention that the main difference between the
heterojunctions A and B is whether the incident electron is in the
middle region or in the side regions. However, since the CBM of
MoS$_2$ is lower than that of the WS$_2$, it is assumed that in
both cases the incident electron is in the MoS$_2$ region, such
that the electron in conduction band of MoS$_2$ can either enter
into the WS$_2$ region or be reflected back to the MoS$_2$,
depending on its energy.

\subsection{$T$ and GH shift in type-A heterojunction}
We consider an electron beam incident on the interface at $x=d$,
from region 2 into region 3, as shown in Fig. 1(a). The electron
wave functions in regions 2 and 3 can be written in terms of
incident, reflected, and transmitted waves as
\begin{eqnarray}\label{psi2}
\psi_{2}(x,y)=\frac{1}{B_2}\left(\begin{array}{cc} A_2\\ \tau
a_2t_2k_2\mathrm{e}^{i\tau\theta_2}\\ \nonumber
\end{array}\right)\mathrm{e}^{i(k_{2x}x+k_yy)}\\
+\frac{r^\tau_{s_z}}{B_2}\left(\begin{array}{c}
A_2\\ -\tau a_2t_2k_2\mathrm{e}^{-i\tau\theta_2}\\
\end{array}\right)\mathrm{e}^{i(-k_{2x}x+k_yy)}\  ,
\end{eqnarray}
and
\begin{eqnarray}\label{psi3}
\psi_{3}(x,y)&=&\frac{t_{s_z}^\tau}{B_3}\left(\begin{array}{c}
A_3\\ \tau a_3t_3k_3\mathrm{e}^{i\tau\theta_3}\\
\end{array}\right)\mathrm{e}^{i(k_{3x}x+k_yy)}\  ,
\end{eqnarray}
where $k_y$ is the same in all regions due to the translational
invariance in $y$ direction, $A_2$, $B_2$, $A_3$, and $B_3$ are
given in Eq. (\ref{h2}), and the coefficients $r^\tau_{s_z}$ and
$t^\tau_{s_z}$ are reflection and transmission coefficients,
respectively, which can be obtained by matching the wave functions
at $x=d$. The result for the reflection coefficient is
\begin{equation}\label{r}
r^\tau_{s_z}=\frac{s_1s_2\sqrt{\frac{F_c}{F_v}}\mathrm{e}^{i\tau
\theta_2}-\mathrm{e}^{i\tau
\theta_3}}{s_1s_2\sqrt{\frac{F_c}{F_v}}\mathrm{e}^{-i\tau
\theta_2}+ \mathrm{e}^{i\tau \theta_3}}\mathrm{e}^{2ik_{2x}d}\ ,
\end{equation}
where $s_1=\mathrm{sgn}(A_1)=\mathrm{sgn}(A_3)$,
$s_2=\mathrm{sgn}(A_2)$, $F_v=\frac{E-E_{2v}-\tau
s_z\lambda_2+h_2s_z}{E-E_{1v}-\tau s_z\lambda_1+h_1s_z}$, and
$F_c=\frac{E-E_{2c}+h_2s_z}{E-E_{1c}+h_1s_z}$. The critical angle
$\theta_c$ for total internal reflection at the interface is given
by
\begin{equation}\label{phic}
\theta_c=\arcsin(\frac{k_3}{k_2}) .
\end{equation}
In the case of $\theta_2>\theta_c$ the electron wave number
$k_{3x}=\sqrt{k_3^2-k_y^2}$ becomes imaginary, leading to an
evanescent wave in region 3, and consequently the electron beam
undergoes a total reflection from the interface at $x=d$. Using
$T=1-|r^\tau_{s_z}|^2$ and Eq. (\ref{r}), the transmission
probability along the $x$-axis can be expressed as
\[T=\left\lbrace
\begin{array}{cl}
\frac{4\sqrt{\frac{F_c}{F_v}}\cos\theta_2\cos\theta_3} {1 +
\frac{F_c}{F_v}+2s_1s_2\sqrt{\frac{F_c}{F_v}}\cos(\theta_2+\theta_3)}
&\text{if   $\theta_2<\theta_c$},\\
0 & \text{if   $\theta_2>\theta_c$}.
\end{array}
\right. \]

The \textit{total} reflection can lead to electron confinement in
region 2 between the two interfaces associated with multiple
reflections from the interfaces at $x=0$ and $x=d$. Furthermore,
based on the stationary phase method \cite{Ghadiri}, the GH
lateral shift of the reflected beam can be obtained as
\begin{equation}\label{sigmare1}
\sigma^\tau_{re,{s_z}}=-\dot{\Phi}_{r^\tau_{s_z}}+2\frac{\tau
{a_2}^2{t_2}^2 {k_2}^2}{{B_2}^2}\dot{\theta}_2+2\dot{k}_{2x}d\  ,
\end{equation}
where $\Phi_{r^\tau_{s_z}}$ is the phase of reflection coefficient
and the dot indicates the derivative with respect to $k_y$. By
calculating $\Phi_{r^\tau_{s_z}}$ from Eq. (\ref{r}) and
substituting it into Eq. (\ref{sigmare1}), after some algebra, we
can express the GH shift of reflected electrons as
\begin{equation}\label{sigmare2}
\begin{array}{c}
\sigma^\tau_{re,{s_z}}=\frac{2s_1s_2k_1(\kappa+\tau
k_y)\sqrt{\frac{F_c}{F_v}}(\frac{\tau
\cos\theta_2}{\kappa}+\frac{\tan\theta_2}{k_2})-2\tau
k_1^2\frac{F_c}{F_v k_{2x}}}{k_1^2\frac{F_c}{F_v}+(\kappa+\tau
k_y)(\kappa+\tau k_y -2\tau
s_1s_2k_1\sqrt{\frac{F_c}{F_v}}\sin\theta_2)}\\
+\frac{2\tau}{(1+\frac{E-E_{2v}-\tau
s_z\lambda_2+h_2s_z}{E-E_{2c}+h_2s_z})k_{2x}} \  ,
\end{array}
\end{equation}
where $\kappa=ik_{3x}$. The obtained
$\sigma^\tau_{re,{s_z}}$ has the order of magnitude of the Fermi
wavelength, $\lambda_F$, (see Ref. \cite{Beenakker} and
\cite{Sun}) which probably impedes its direct measurements.
However, when the total internal reflection occurs, due to the
multiple reflection from the interfaces in region 2, the lateral
shifts of the reflected beams along the interface will accumulate
and considerably exceed from $\lambda_F$, after the beams travel a
sufficiently long distance inside the region.

Therefore, if the incidence angle exceeds the critical angle, an
electron waveguide forms in the type-A heterojunctions, in which
the electrons with quasibound states are confined in $x$
direction, while they propagate in $y$ direction. Note that the
energy spectrum of these bound states can be calculated by
matching the propagating wave in region 2 with the evanescent
waves in regions 1 and 3 at the interfaces $x=0$ and $x=d$,
respectively (see Ref. \cite{Beenakker} and \cite{Sun}).

\subsection{$T$ and GHL shift in type-B heterojunction}
We now consider the type-B heterojunction in which the electrons
injected from region 1 into region 2 propagate across the
interfaces at $x=0$ and $x=d$, as shown in Fig. 1(b). In this
case, the wave function of electron in each region can be written
as
\begin{eqnarray}\label{psi1}
\begin{array}{c}
\psi_{1}(x,y)=\frac{1}{B_1}\left(\begin{array}{c}
A_1\\ \tau a_1t_1k_1\mathrm{e}^{i\tau\theta_1}\\
\end{array}\right)\mathrm{e}^{i(k_{1x}x+k_yy)}\\
+\frac{r^\tau_{s_z}}{B_1}\left(\begin{array}{c}
A_1\\ -\tau a_1t_1k_1\mathrm{e}^{-i\tau\theta_1}\\
\end{array}\right)\mathrm{e}^{i(-k_{1x}x+k_yy)}\ ,
\end{array}
\end{eqnarray}
\begin{eqnarray}\label{psi2IIb}
\begin{array}{c}
\psi_{2}(x,y)=\frac{\alpha}{B_2}\left(\begin{array}{c}
A_2\\ \tau a_2t_2k_2\mathrm{e}^{i\tau\theta_2}\\
\end{array}\right)\mathrm{e}^{i(k_{2x}x+k_yy)}\\
+\frac{\beta}{B_2}\left(\begin{array}{c}
A_2\\ -\tau a_2t_2k_2\mathrm{e}^{-i\tau\theta_2}\\
\end{array}\right)\mathrm{e}^{i(-k_{2x}x+k_yy)}\  ,
\end{array}
\end{eqnarray}
\begin{eqnarray}\label{psi3IIb}
\psi_{3}&=&\frac{t_{s_z}^\tau}{B_3}\left(\begin{array}{c}
A_3\\ \tau a_3t_3k_3\mathrm{e}^{i\tau\theta_3}\\
\end{array}\right)\mathrm{e}^{i(k_{3x}x+k_yy)}\  ,
\end{eqnarray}
where $a_1=a_3$, $t_1=t_3$, $k_1=k_3$, $\theta_1=\theta_3$,
$A_1=A_3$, and $B_1=B_3$. Note that the coefficients $A_1$, $B_1$,
$A_2$, and $B_2$ are determined from Eq. (\ref{h2}). The critical
angle for total reflection in this case is given by
\begin{equation}\label{phic2}
\theta_c=\arcsin(\frac{k_2}{k_1})\  .
\end{equation}

When $\theta_1<\theta_c$, the electrons can be partially
transmitted through the proposed heterojunction. The coefficients
$r^\tau_{s_z}, \alpha, \beta$ and $t_{s_z}^\tau$ are obtained by
matching wave functions at the interfaces $x=0$ and $x=d$. The
result for the transmission coefficient, $t_{s_z}^\tau$, is
\begin{equation}\label{trIIb}
t^\tau_{s_z}=\frac{-2s_1s_2\sqrt{\frac{F_v}{F_c}}\cos\theta_1\cos\theta_2
\mathrm{e}^{-ik_{1x}d}}{-2s_1s_2\sqrt{\frac{F_v}{F_c}}~\cos\theta_1\cos\theta_2\cos(k_{2x}d)+iD}\
,
\end{equation}
where
$D=[(1+\frac{F_v}{F_c})-2s_1s_2\sqrt{\frac{F_v}{F_c}}\sin\theta_1\sin\theta_2]
\sin(k_{2x}d)$. Here, $T=|t^\tau_{s_z}|^2$ and using the
derivative of the phase of transmission coefficient with respect
to $k_y$, one can write the GHL shift of the transmitted electrons
as \cite{Ghadiri}
\begin{equation}\label{GHLt-endIIb}
\sigma^{\tau}_{tr,s_z}=\frac{[(8+2\frac{k^2_0}{k^2_{1x}}+2\frac{k^2_0}{k^2_{2x}})
\frac{\sin(2k_{2x}d)}{2k_{2x}d}-2\frac{k^2_0}{k^2_{2x}}]d\tan\theta_1}
{4\cos^2(k_{2x}d)+\frac{k_0^4}{k^2_{1x}k^2_{2x}}\sin^2(k_{2x}d)}\
,
\end{equation}
where $k^2_0=-s_1s_2\alpha k_1k_2+2k^2_y$ and
$\alpha=\sqrt{F_c/F_v}(1+F_v/F_c)$.

In the vicinity of resonance positions the lateral shift can be
greatly enhanced (see Refs. \cite{ChenEPJB,Ghadiri}). At resonance
positions which are determined by the condition $k_{2x}d=n\pi$,
$(n=1, 2,3,....)$, the heterojunction becomes transparent ($T=1$)
and $\sigma^{\tau}_{tr,s_z}$ acquires local maxima which are
obtained from
\begin{equation}\label{max}
\left.\sigma^{\tau}_{tr,s_z}\right|_{k_{2x}d=n\pi}=
\frac{k^2_{0}d\tan\theta_1}{2k^2_{2x}}=\left.n\sigma^{\tau}_{tr,s_z}\right|_{k_{2x}d=\pi}
.
\end{equation}
Note that in the case of $\theta_1>\theta_c$, the transmission of
electrons become negligible and $\sigma^{\tau}_{tr,s_z}$ will be
of the order of $\lambda_F$ as in the type-A heterojunction. Here,
in contrast to Sec. II.A, the reflection and transmission
coefficients are obtained by applying the boundary conditions on
both interfaces at $x=0$ and $x=d$. Therefore, the transmission
probabilities and lateral shifts depend on $d$. It is worth
mentioning that in contrast to optical beams in 2D materials such
as graphene and single-layer boron-nitride \cite{OpticsLetters},
where the GH shift does not depend primarily on the wavelength of
the incident light beam, Eqs. (\ref{sigmare2}) and
(\ref{GHLt-endIIb}) for lateral shifts of electron beams depend on
the Fermi energy (electron Fermi wavelength) via Eq. (\ref{Ek}).

\section{results and discussion}
In order to study the effect of band alignment of TMDs on
ballistic transport and lateral shift of electron beams with
different flavors in the planar heterojunctions, we first show in
Fig. 1(c)-(e) the valence band maximums (VBMs) and conduction band
minimums (CBMs) at the MoS$_2$/WS$_2$ interface for the cases with
normal and ferromagnetic WS$_2$ regions. Here, a flavor is denoted
as $(s_z,\tau)$, which represents an electron with spin $s_z$ in
valley $\tau$. Therefore, there are four different flavors as
($\pm 1$,K) and ($\pm 1$,K$'$). Parameters $a$, $t$, $\lambda$,
and $E_{c(v)}$ in each TMD material are chosen according to Refs.
\cite{Xiao} and \cite{Kang}, respectively, where the energy band
edges are measured with respect to the vacuum level (zero point
energy).

In the normal regions, the CBMs of all flavors are the same but
the VBMs are partially split as a result of the coupled spin and
valley degrees of freedom in TMD monolayers (see Eq. \ref{Ek}). In
fact the flavors with the same amount of $s_z\tau$ have the same
VBM as shown in Fig. 1(c). In the presence of magnetic proximity
effect, however, the spin degeneracy at the conduction-band edges
is lifted and the spin splitting in the VBM becomes strongly
valley dependent as can be seen in Fig. 1(d) and (e).
\begin{figure}
\center\includegraphics[width=0.85\linewidth]{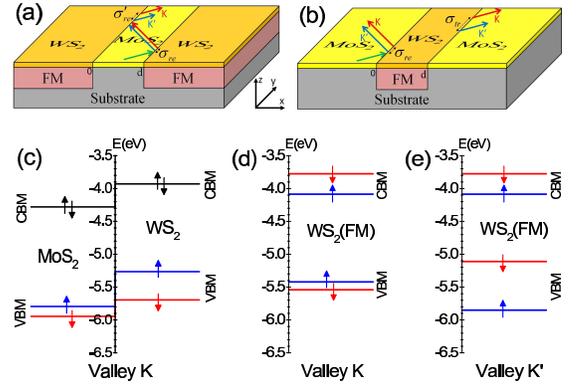}
\caption{(Color online) Schematic models for (a)
WS$_2$/MoS$_2$/WS$_2$ (type-A) and (b) MoS$_2$/WS$_2$/MoS$_2$
(type-B) lateral junctions. The WS$_2$ region is placed in close
proximity to a ferromagnetic (FM) substrate. (c)-(e) Band
alignments at MoS$_2$/WS$_2$ interface for the cases with (c)
normal and ((d),(e)) ferromagnetic WS$_2$ regions in which
$h$=0.155 eV [shown as WS$_2$(FM)]. In (a) and (b), the green
arrow shows the initial incident beam, whereas the red and blue
ones indicate the separated electrons in valleys K and K$'$,
respectively. The arrows in (c)-(e) show the spin of electrons in
the VBM and CBM. Note that the spin directions in (c) are inverted
for K$'$ valley electrons. Also, the energy-band edges of MoS$_2$
region in (d) and (e) are the same as those in (c) (not shown).
The VBMs and CBMs are measured with respect to the vacuum level
\cite{Kang}.}
\end{figure}

According to Eqs. (2), (7), and (13), the value of critical angle
depends on the electron energy $E$, spin, and valley indexes. To
show this, the critical angle is depicted in terms of $E$ in Figs.
2(a) and (b) for the four flavors, when electrons are incident
from MoS$_2$ region on normal WS$_2$ and ferromagnetic WS$_2$
regions, respectively. As can be seen in Fig. 2(a), the flavors
with the same values of $s_z\tau$ have the same critical angles
due to the absence of magnetic proximity effect. In the presence
of exchange field, however, the spin-valley symmetry is partially
(fully) broken and the electrons in the conduction (valence) band
of WS$_2$ region are separated into two (four) different band edge
energies and critical angles, as shown in Fig. 2(b). There are two
different flavors in the CBM (at $k_2=0$) due to the lack of
spin-orbit splitting in the conduction band and hence the four
flavors become doubly degenerate as are seen in Fig. 1(d) and (e).
This means that the critical angles are doubly degenerate at
energy values of the CBMs. To explain the behavior of $\theta_c$
in terms of energy, it should be mentioned that the critical angle
is zero when the electron energy is lower than the CBM of each
flavor (energy gap region). For a given energy in the conduction
band of WS$_2$, however, the transverse wave vector $k_y$ which is
conserved, increases as the incidence angle increases and exceeds
the Fermi wave vector $k_F(\equiv k)$. As a result, the
longitudinal wave vector $k_x$ becomes imaginary, and hence, the
incidence angle reaches its critical value. Moreover, for a given
energy and an incidence angle, all flavors have almost the same
$k_y$ values due to the relatively small spin-orbit coupling and
absence of exchange field in MoS$_2$ region, whereas the flavors
have different $k_F$ values in WS$_2$ region, and consequently,
different critical angles are obtained, as shown in Fig. 2(b).
\begin{figure}
\center\includegraphics[width=0.65\linewidth]{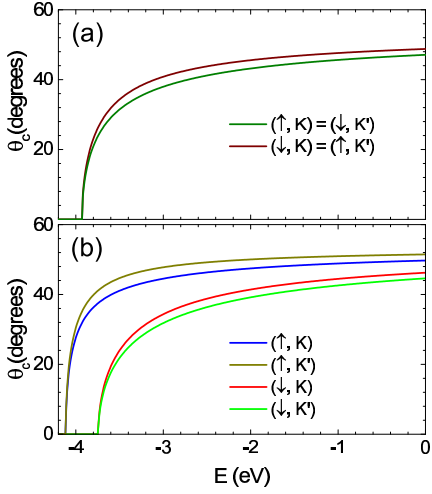}
\caption{(Color online) Dependence of critical angle on incident
electron energy at the MoS$_2$/WS$_2$ interface for the cases with
(a) normal and (b) ferromagnetic WS$_2$ regions in which $h$=0.155
eV.}
\end{figure}

From Figs. 1(c)-(e), one can consider the three energy intervals;
(i) $E>E_{c(\mathrm{WS}_2)}+h$, (ii)
$E_{c(\mathrm{WS}_2)}-h<E<E_{c(\mathrm{WS}_2)}+h$, and (iii)
$E_{c(\mathrm{MoS}_2)}<E<E_{c(\mathrm{WS}_2)}-h$. Moreover, in the
absence of spin-orbit coupling, the energy gap in each region is
determined by $\Delta_j=E_{jc}-E_{jv}$. Therefore, any errors in
$E_{jc}$ and $E_{jv}$ values can quantitatively affect the band
gaps and the numerical results. However, this effect can be
compensated by choosing an appropriate exchange field $h$ so that
the electrons with energy $E$ lie within one of the energy
intervals (i)-(iii). Each flavor, injected from MoS$_2$ region can
enter into the WS$_2$ region, if the Fermi energy is higher than
the CBM of that flavor and simultaneously, the incidence angle of
electrons is less than the critical angle of that flavor (see Fig.
2(b)). This means that in type-A heterojunction, the electron
leaves the waveguide (middle region), while in type-B
heterojunction the electron propagates through the structure. In
the following, we present our numerical results of transmission
probabilities and lateral shifts in both heterojunctions, using
different parameters.

Fig. 3(a) shows the transmission probability in terms of incidence
angle $\theta_2$ for the typical low lying energy -3.70 eV and
$h_1=h_3=0.1\Delta_{\mathrm{WS}_2}=0.155$ eV in type-A
heterojunction. Since this energy value lies in the conduction
band of all flavors (energy interval (i)), the incident electrons
belonging to each flavor can propagate into region 3. According to
Fig. 2b different flavors of electrons have different critical
angles that can also be seen in Fig. 3(a). With increasing
$\theta_2$, the transmission probability drops to zero sharply as
$\theta_2$ approaches to critical angle of each flavor. The
different $T$ values for spin-up and spin-down electrons are
mostly related to the spin splitting of CBMs in the ferromagnetic
WS$_2$ region. The incident spin-up electrons at the
MoS$_2$/WS$_2$(FM) interface have higher probability to propagate
into region 3 because the occupation of spin-up levels is higher
than that of the spin-down levels. The small difference in $T$
values for the flavors with the same spin is due to the spin-orbit
coupling in both materials, particularly in WS$_2$ region. The
corresponding GH lateral shifts are depicted in Fig. 3(b). The
abrupt increase in the GH shift of each flavor is related to the
corresponding critical angle. When the incidence angle $\theta_2$
is less than $\theta_c$, the sign of GH shift can be positive or
negative depending on the spin and valley indexes, whereas the
lateral shift of all flavors is pure positive when
$\theta>\theta_c$. The difference between GH shift values of
different flavors is a consequence of their different band offsets
in MoS$_2$/WS$_2$ heterojunctions.
\begin{figure}
\center\includegraphics[width=0.95\linewidth]{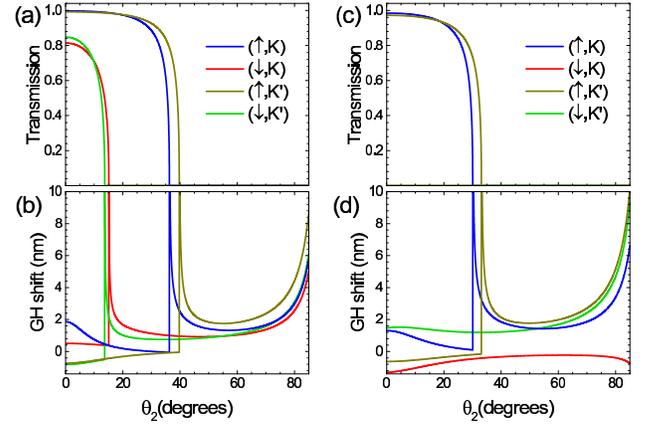}
\caption{(Color online) Transmission probability and GH shift of
the reflected electrons with ((a) and (b)) $E=-3.70$ eV and ((c)
and (d)) $E=-3.90$ eV as a function of incidence angle $\theta_2$
in type-A heterojunction with $h_1=h_3=0.155$ eV.}
\end{figure}

If we choose
$\theta_c(-1,\mathrm{K}')<\theta_2<\theta_c(-1,\mathrm{K})$, only
spin-down electrons in K$'$ valley are totally reflected, and
hence, they propagate in region 2 after undergoing a consecutive
total reflection from parallel interfaces at $x=0$ and $x=d$. In
such a case, the other three flavors can penetrate into the
regions 1 and 3, and eventually they disappear from region 2 after
consecutive reflections from the interfaces, suggesting a
spin-valley polarized beam inside the channel in region 2. For
$\theta_c(-1,\mathrm{K})<\theta_2< \theta_c(1,\mathrm{K})$, on the
other hand, only spin-down electrons are allowed to propagate
inside the channel, whereas the spin-up electrons leave the
ferromagnetic WS$_2$ region. Since each flavor experiences a
different GH shift value (see Fig. 3(b)), the spin-down electrons
can be well-separated inside the channel after traveling a
sufficiently long distance. If $\theta_2$ is chosen greater than
the critical angles of the four flavors, then all electrons will
be totally reflected into the region 2, and after traveling a
sufficiently long distance inside the channel, the four electron
beams with different spin-valley indexes can be spatially
separated.

To see how the electrons with a different energy may affect the
result, we have also depicted $T$ and GH shift of the reflected
electrons at $E=-3.9$ eV in Figs. 3(c) and (d), respectively.
Since this energy value lies below the CBM of the spin-down
electrons (energy interval (ii)), the propagation of these
electrons is blocked ($T=0$) regardless of their incidence angle
value. As a result, for the corresponding flavors, the former
abrupt increase in the GH shift values in Fig. 3(d) does not
exist. The spin-up electrons, however, can propagate into the
channel or travel outside in regions 1 and 3, depending on whether
or not the incidence angle exceeds the corresponding critical
angle. Accordingly, depending on the value of incidence angle, we
can expect two, three or four well-separated flavors inside the
waveguide channel. If the energy of electrons lies between the CBM
of spin-up flavors in the ferromagnetic WS$_2$ region and the CBM
of MoS$_2$ region (energy interval (iii)), all flavors will be
totally reflected inside the WS$_2$ region. From the lateral shift
values (not shown here) we found that all flavors were
well-separated after passing a sufficiently long distance inside
the channel. Therefor, in such a case the heterojunction acts as a
fully spin-valley beam splitter regardless of the incidence
angles.
\begin{figure}
\center\includegraphics[width=0.95\linewidth]{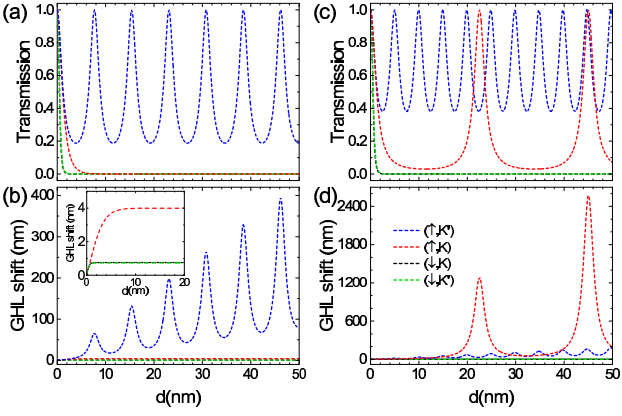}
\caption{(Color online) Dependence of (a, c) transmission
probability and (b, d) GHL shift of electrons on the width $d$ of
WS$_2$ region in type-B heterojunction. The parameters are
$E=-3.85$ eV and $h_2=0.155$ eV. The incidence angles in (a, b)
and (c, d) are $\theta_1=34^\circ$ and $\theta_1=32^\circ$,
respectively. The inset shows how the three flavors saturate to
constant values as $d$ increases. Note that the legends in (a)-(c)
are the same as those in (d).}
\end{figure}

We continue by presenting numerical results for type-B lateral
heterojunction in which the electron beams are incident from
region 1 on MoS$_2$/WS$_2$ interface at $x=0$, as shown in Fig.
1(b). Here, unlike the results of type-A heterojunction, there is
a possibility of constructive interference of forward and backward
moving waves in the middle region which manifests itself as a
perfect transmission and also considerable increase in the GHL
shift value. Fig. 4(a) shows the plots of $T$ as a function of
width $d$ of the ferromagnetic WS$_2$ region for low energy
electrons with $E=-3.85$ eV. As can be seen, $T$ has resonant
features for propagating flavor (1,K$'$), while it is evanescent
for the other three flavors which decay exponentially with
increasing $d$. This behavior originates from spin splitting of
the conduction band edge in region 2, resulting from exchange
field and also different critical angles for different flavors,
and can be understood from Figs. 1(d)-(e) and 2(b). Since the
chosen Fermi energy lies inside the spin-down energy gap of the
middle region (energy interval (ii)), evanescent modes appear for
this type of electrons. For small $d$ values, the incident
spin-down electrons can tunnel through the corresponding energy
gap and propagate in region 3, i.e., $T$ is nonzero, whereas with
increasing $d$, $T$ decays exponentially to zero. Although both
spin-up flavors lie energetically in the conduction band of middle
region, since incidence angle of electrons is chosen as
$\theta_c(1,\mathrm{K})<\theta_1=34^{\circ}<\theta_c(1,\mathrm{K}')$
(see Fig. 2(b)), $k_{2x}$ becomes imaginary for the flavor (1,K)
and consequently this flavor also finds evanescent character,
decaying exponentially with increasing $d$. Moreover, since
Im$(k_{2x})$ for electrons with spin-down flavor is greater than
that for electrons with flavor (1,K), $T$ decays more rapidly for
spin down flavors compared to that for flavor (1,K). The
corresponding GHL shift of the transmitted electrons,
$\sigma^{\tau}_{tr,s_z}$, as a function of $d$ is depicted in Fig.
4(b). For evanescent states, the GHL shift is of the order of
$\lambda_F$ and saturates to a constant value (see the inset in
Fig. 4(b)). For propagating states (flavor (1,K$'$)), however, the
GHL shift oscillates with $d$ and demonstrates a resonant
character, similar to the corresponding transmission in Fig. 4(a).
We can see that at resonance positions the structure is fully
transparent ($T=1$) and GHL shift has local maxima which are given
by Eq. (\ref{max}).
\begin{figure}
\center\includegraphics[width=0.95\linewidth]{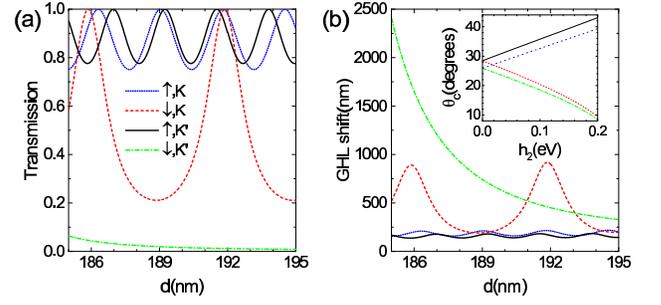}
\caption{(Color online) Dependence of (a) transmission probability
and (b) GHL shift of electrons with different flavors on the width
$d$ of the WS$_2$ region in type-B heterojunction. The parameters
are $E=-3.70$ eV, $\theta_1=20.8^\circ$, and $h_2=0.07$ eV. The
inset in (b) shows $\theta_c$ as a function of exchange field
$h_2$.}
\end{figure}

For a given $E$ and $\theta_1$, the quantities $k_1$, $k_2$,
$k_y$, and $k_{2x}$ have fix values and consequently the resonance
positions $d_n$ and the corresponding GHL shift values will be
proportional to $n$ (number of resonances), as can be seen in Fig.
4(b) and (d). We should note that by flipping the direction of
exchange field the junction can filter the flavors (1,K),
(1,K$'$), (-1,K$'$) and only the flavor (-1,K) will be allowed to
pass through the system. In Fig. 4(c) and (d) the incidence angle
is taken as $\theta_1=32^\circ$ which is less than the critical
angle of spin-up flavors (see Fig. 2(b)). Therefore, both spin-up
flavors contribute in transmission and exhibit different behaviors
in their $T$ and GHL shifts, due to a difference in their $k_{2x}$
values. In reality, $k_{2x}$ for electrons with flavor (1,K) is
smaller than that for electrons with flavor (1,K$'$), and hence,
the corresponding period and amplitude of oscillations in both $T$
and GHL shift curves are larger compared to those for the flavor
(1,K$'$). As a result, for some $d$ values a large spatial
separation between the two flavors as large as longitudinal width
of the incident beam \cite {Song} can be seen in Fig. 4(d),
suggesting that the lateral heterojunction with the given
parameters can act as a valley beam splitter for spin-up electrons
and simultaneously block the spin-down electrons, acting as a spin
filter. This spin selection for electron propagation can be
flipped by reversing the direction of exchange field.
\begin{figure}
\center\includegraphics[width=0.85\linewidth]{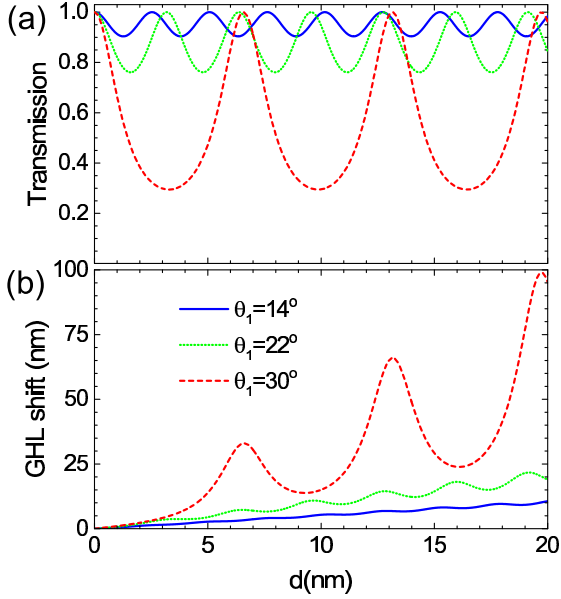}
\caption{(Color online) Dependence of (a) transmission probability
and (b) GHL shift of transmitted electrons with flavor (1,K) on
the width $d$ of WS$_2$ region in type-B heterojunction for three
different incidence angles. The parameters $E$ and $h_2$ are the
same as those in Fig. 4.}
\end{figure}

In order to have propagating states for all flavors, electron
energy should be located in the conduction band of the flavors
(energy interval (i)) and the incidence angle should also be less
than the corresponding critical angles. Due to the relatively
\textit{large} difference between critical angles of the two
spin-up flavors and those of the two spin-down flavors, also due
to the relatively \textit{small} difference between the critical
angles of the two spin-up flavors (Fig. 2(b)), the difference
between incidence angle and the critical angle for the two spin-up
flavors increases and as we will show later, this leads to a
considerable reduction in their GHL shift values, and hence, small
valley splitting. To overcome this issue one can reduce the
magnetic proximity effect that is equivalent to a reduction in
$h_2$ value. For such a case, in Fig. 5(a) and (b) we have shown
$T$ and GHL shift as a function of $d$ for electrons with $E=-3.7$
eV and $\theta_1=20.8^\circ$ in the presence of the relatively
weak exchange field $h_2$=0.07 eV. Moreover, the inset in Fig.
5(b) shows how the critical angle of each flavor in the
ferromagnetic WS$_2$ region is affected by the exchange field
$h_2$. From Figs 5(a) and 5(b), it is evident that all flavors can
pass through the device with an oscillatory behavior in their $T$
and GHL shift curves. Interestingly, for type-B heterojunctions
with $d\sim186$ nm, the difference between lateral shift of each
flavor and that of the other flavors increases considerably and
therefore the heterojunction acts as a fully spin-valley beam
splitter.
\begin{figure}
\center\includegraphics[width=0.8\linewidth]{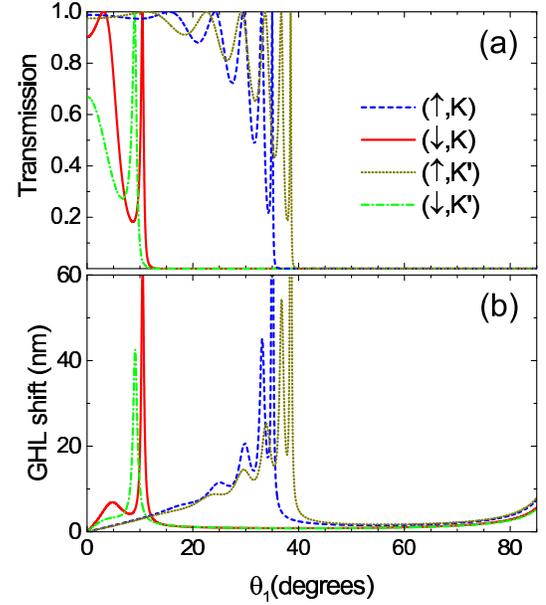}
\caption{(color online) Dependence of (a) transmission probability
and (b) GHL shift of the transmitted electrons on the incidence
angle $\theta_1$ in type-B heterojunction. The parameters are
$d$=10 nm, $E=-3.73$ eV, and $h_2=0.155$ eV.}
\end{figure}

To show how the propagating states (see, for instance, Fig. 4) are
affected by changing $\theta_1$, we have depicted in Fig. 6(a) and
(b) the transmission $T$ and GHL shift of electrons in terms of
$d$ for propagating flavor (1,K) at different angles
$\theta_1<\theta_c$. We can see that the GHL shift is considerably
small when $\theta_1\ll\theta_c$. According to the resonance
condition $k_{2x}d=n\pi$ and Eq. (\ref{max}), the distance between
resonance positions and the amplitude of oscillations in both $T$
and GHL shifts increases with increasing $\theta_1$. In fact, as
$\theta_1$ approaches $\theta_c$, $k_{2x}\rightarrow 0$ and
consequently $d$ and the lateral shift values at which resonances
occur can increase dramatically, indicating that valley transport
in MoS$_2$/WS$_2$ heterojunctions can be controlled by incidence
angle of electrons.

Moreover, for a fixed $d$ value, the functionality of type-B
heterojunctions can be explored by changing the incidence angle of
electrons with different flavors. In Fig. 7, several resonances in
$T$ and lateral shift can be seen which are obtained from the
relation $\theta_{1,n}=\arcsin(\sqrt{k_2^2-n^2\pi^2/d^2}/k_1)$. It
is evident that when $\theta_1$ reaches the critical angle of each
flavor the corresponding $T$ value drops to almost zero, resulting
a considerable reduction in the GHL shift (see also Fig. 4(a),
(b), and the inset). For the case of $\theta_1<\theta_c$(-1,K$'$),
all flavors pass through the structure (see Fig. 7(a)), but as
discussed earlier, the difference between GHL shift values for
spin-up flavors is negligible, due to a considerable difference
between the value of $\theta_1$ and $\theta_c$ for spin-up flavors
(see Fig. 7(b)). When $\theta_c(-1,$K$)<\theta_1<\theta_c(1,$K),
spin-down flavors are almost blocked, whereas spin-up flavors pass
through the structure and a considerable difference between GHL
shifts of spin-up flavors is obtained before $\theta_1$ reaches
$\theta_c(1,$K). This means that the heterojunction can
effectively split the flavors when the electron beams enter into
the region 3. Note that for the case of
$\theta_c(1,$K$)<\theta_1<\theta_c(1,$K$')$, only the electrons
with flavor (1,K$'$) can pass through the structure, suggesting
MoS$_2$/WS$_2$ heterojunction as promising structures for
spin-valley filtering.

Note that although the measurement of electric GH shifts in 2D
materials is still an open challenge due to the electron
scattering, smallness of GH shifts in experiments, and difficulty
in preparation of a well-collimated electron beam
\cite{ChenOptics}, the present findings can improve our
fundamental understanding of electronic version of GH effect and
also provide a new platform for application of TMD heterojunctions
as spin-valley beam filters and/or splitters.

\section{conclusions}
In summary, we have explored theoretically the effect of band
alignments on spin-valley transport and lateral shift of electrons
in MoS$_2$/WS$_2$ planar heterojunctions in which the WS$_2$
region is placed in close proximity to a ferromagnetic substrate.
We found that electron waveguiding can occur in
WS$_2$/MoS$_2$/WS$_2$ heterojunction for propagating electrons
inside the MoS$_2$ monolayer due to the electron confinement in
the central region. In MoS$_2$/WS$_2$/MoS$_2$ heterojunction,
however, transmission resonances formed in the WS$_2$ region play
the main role in generation of strong lateral displacements of
electron beam transmitted through the structure. In both
heterojunctions, the lateral shift of electrons induced by band
alignments of the two constituent TMD monolayers is spin and
valley dependent. It is shown that in these heterojunctions,
electrons with distinct spin and valley can be filtered and/or
spatially well-separated by tuning the Fermi energy and incidence
angle of electrons. Our findings suggest new generation of
nanodevices based on lateral TMD heterojunctions which can produce
fully spin-valley polarized currents without external electrical
tuning.

\section*{Acknowledgement}
This work is partially supported by Iran Science Elites Federation
(11/66332).

\end{document}